\pdfoutput=1
\documentclass[10pt,conference,letterpaper]{IEEEtran}
\usepackage{amsmath,amssymb,bm}
\usepackage{cite}
\usepackage{graphicx}
\usepackage{booktabs}
\usepackage{url}
\usepackage{eso-pic}
\AddToShipoutPictureFG*{\AtPageLowerLeft{\put(0,\LenToUnit{0.35in}){\makebox[\paperwidth]{\parbox{0.8\paperwidth}{\centering\footnotesize
This work has been submitted to the IEEE for possible publication. Copyright may be transferred without notice, after which this version may no longer be accessible.}}}}}
\graphicspath{{./}}
\title{First-Frame Guided Target Propagation for Vision-Radar Beam Prediction in Crowded V2I}

\author{\IEEEauthorblockN{Junyi Chen$^{*\dagger}$, Zhaohui Huang$^{*}$, Pengyu Wang$^{*}$, Xiao Liang$^{\dagger\ddagger}$, Tianqi Mao$^{\dagger}$, and Zhaocheng Wang$^{*\dagger}$}
\IEEEauthorblockA{$^{*}$Department of Electronic Engineering, Tsinghua University, Beijing 100084, China\\
$^{\dagger}$State Key Laboratory of Environment Characteristics and Effects for Near-Space, Beijing 100081, China\\
$^{\ddagger}$Academy of Advanced Interdisciplinary Research, Beijing Institute of Technology, Beijing 100081, China\\
Emails: 
junyi-ch23@mails.tsinghua.edu.cn, hzh21@mails.tsinghua.edu.cn, wangpengyu.2023@tsinghua.org.cn,\\
liang.xiao@bit.edu.cn, maotq@bit.edu.cn, zcwang@tsinghua.edu.cn}}

\begin{document}
\maketitle

\begin{abstract}

Beam prediction can reduce training overhead for millimeter-wave (mmWave) vehicle-to-infrastructure (V2I) links, but in crowded roadside scenes it is difficult to associate camera and radar evidence with the target vehicle. In this paper, we propose a target-conditioned vision-radar beam prediction framework. First, a first-frame guided target propagation scheme identifies the target vehicle once and propagates this association through the image sequence, including occluded frames. Then, a calibrated image-to-range mapping transfers the propagated box to the radar domain, where it selects local range-angle/range-velocity (RA/RV) windows of the same vehicle. The target masks and local RA/RV windows are fused by a multimodal network for beam prediction. Experiments on DeepSense 6G Scenario 31 show that the proposed propagation misses the target in $0.6\%$ of the test frames and $5.3\%$ of the occluded frames. Consequently, the framework improves distance-based accuracy (DBA) by $0.0358$ and Top-1 accuracy by $3.97\%$ over the scene-level baseline and outperforms three image--radar fusion networks.

\end{abstract}

\section{Introduction}

Millimeter-wave (mmWave) communication systems can support high data rates through directional beamforming, but beam acquisition and maintenance still incur substantial overhead in mobile vehicle-to-infrastructure (V2I) scenarios \cite{Rappaport2019}. This has motivated sensing-aided beam prediction, which uses side information to reduce beam training overhead.

Early learning-based beam prediction relied on communication-side observations and temporal correlations to reduce exhaustive beam search \cite{Alkhateeb2018,Wang2025Flexible}, and sub-6 GHz side information has been used to infer mmWave beams without exhaustive training at the target band \cite{Alrabeiah2020Sub6}. These methods, however, rely solely on wireless measurements and provide no explicit geometric information about the user, which can be obtained directly from visual and radar sensing. Camera-based methods use images to estimate user location or predict beam indices in real V2I settings \cite{Charan2022VisionPosition}, radar-aided methods exploit range, Doppler, and motion cues \cite{Demirhan2022Radar}, and multimodal approaches combine their complementary appearance and geometric evidence \cite{Nie2023VisionRadar,Tian2023TII,Zhu2025CMDF}. Semantic methods show that target-level boxes or masks can outperform raw scene images \cite{Imran2023SemanticDemo}, and binary target masks have been combined with occlusion-aware training of a sequence-to-sequence beam predictor \cite{Park2025Occlusion}. To select the target, user-identification and multi-candidate methods associate a transmitting user with visual candidates in crowded scenes \cite{Charan2024UserIdentification,Zarei2026CameraAssisted,Charan2025MultiCandidate}, and the user is then followed over time by nearest-neighbor box association \cite{Park2025Occlusion} or multi-object trackers \cite{Li2024VAMUBT}.

In crowded scenes, however, both sensors capture all vehicles within their field of view. Scene-level fusion methods \cite{Nie2023VisionRadar,Tian2023TII,Zhu2025CMDF} encode the entire scene, and the features corresponding to the served user must be inferred implicitly from beam labels. Target-level methods resolve this ambiguity only in the image domain. The radar range bins occupied by the same vehicle depend on camera viewpoint, camera--radar pose, road geometry, and radar discretization; hence, the image-side identity must additionally be mapped to the radar domain. Moreover, nearest-neighbor association and multi-object trackers lose the target after occlusions, when the reappearing, partially visible vehicle is attached to a neighbor or fails to match its previous track.

These gaps need not be closed by re-identifying the target in every frame. Since a vehicle moves continuously, an identity established once in the first frame suffices, as long as it is carried through occlusions in the image domain and transferred to the radar domain. We therefore propagate the target through the image sequence from its first-frame box and project the propagated box onto the radar range axis through a calibrated image-to-range mapping, yielding a target-aligned range-angle/range-velocity (RA/RV) sequence.

The main contributions are summarized as follows.
\begin{itemize}
    \item We propose first-frame guided target propagation, which tracks the target vehicle in the image domain from its first-frame box using motion prediction and appearance-consistent detection, predicts its box from pre-occlusion motion while it is occluded, and recovers it once it reappears.
    \item We develop a calibrated image-to-range mapping that requires no camera--radar registration and transfers the propagated box to local RA/RV windows of the same vehicle, aligning the radar input with the image-side target.
    \item On DeepSense 6G Scenario 31, the proposed propagation misses the target in $5.3\%$ of the occluded frames, versus $20.9\%$ for the best multi-object tracker, and joint visual and radar conditioning raises distance-based accuracy (DBA) from $0.8463$ to $0.8821$ and Top-1 accuracy from $47.69\%$ to $51.66\%$ over the scene-level baseline, exceeding three image--radar fusion networks.
\end{itemize}

\section{System Model and Problem Formulation}
\subsection{System Model}
Consider a roadside V2I system in which a mmWave base station (BS) serves a target vehicle while being equipped with a communication array, an RGB camera, and a radar sensor. The BS employs a predefined beamforming codebook $\mathcal{F}=\{\mathbf{f}_1,\ldots,\mathbf{f}_B\}$, where $B$ is the number of candidate beams and $\mathbf{f}_j$ denotes the $j$th beamforming vector. At time step $t$, the camera captures an RGB frame $\mathcal{I}_t \in \mathbb{R}^{V \times U \times 3}$ of width $U$ and height $V$, and the radar sensor provides synchronized sensing measurements. We denote the preprocessed radar observation at time $t$ by $\mathcal{R}_t$. For downlink transmission, if beam $\mathbf{f}_j$ is used at time $t$, the received signal at the target vehicle can be written as
\begin{equation}
y_t^{(j)}=\mathbf{h}_t^{\mathsf{H}}\mathbf{f}_j s_t+n_t,
\end{equation}
where $\mathbf{h}_t$ denotes the mmWave channel vector between the BS and the target vehicle, $s_t$ is the transmitted symbol, and $n_t$ is additive noise.

\subsection{Motivation and Problem Formulation}
Let $j^{\star}_t \in \{1,\ldots,B\}$ denote the optimal beam index at time $t$. In the conventional formulation, the best beam is the codeword that maximizes the received power, i.e.,
\begin{equation}
j^{\star}_t=\arg\max_{j\in\{1,\ldots,B\}} \left|\mathbf{h}_t^{\mathsf{H}}\mathbf{f}_j\right|^2.
\end{equation}
Since explicit channel estimation and exhaustive beam search incur substantial overhead, sensing-aided beam prediction aims to infer this optimal beam index directly from side information. Let $L$ denote the temporal observation length. A standard supervised model $f_{\Theta}:(\mathcal{I}_{t-L+1:t}, \mathcal{R}_{t-L+1:t}) \rightarrow \hat{j}_{t}$ maps these observations to a predicted beam index $\hat{j}_t$, where $\Theta$ denotes the trainable parameters.

In a crowded scene, $\mathcal{I}_t$ and $\mathcal{R}_t$ contain several vehicles, and only one of them is the served user. If features from a neighboring vehicle dominate, the classifier may predict that vehicle's beam instead of the target vehicle's. We therefore consider a target-conditioned predictor
\begin{equation}
f_{\Theta}:\left(\mathcal{M}_{t-L+1:t}, \widetilde{\mathcal{R}}_{t-L+1:t}\right) \rightarrow \hat{j}_{t},
\label{eq:conditioned_predictor}
\end{equation}
where $\mathcal{M}_t\in\{0,1\}^{256\times256}$ is the resized binary mask of the target vehicle, equal to one on its silhouette and zero elsewhere, and $\widetilde{\mathcal{R}}_t$ retains only the radar returns at its range. Both are derived from a target box $\mathbf b_t$ that is identified once, in the first frame of a continuous segment, and then propagated. The problem thus couples training $f_{\Theta}$ to maximize $\Pr(\hat j_t=j^{\star}_t)$ with two association tasks: propagating the first-frame identity through the image sequence, including occluded frames, and transferring it to the radar maps.

\section{Proposed Methodology}
\subsection{Overview}
\begin{figure*}[t]
\centering
\includegraphics[width=0.88\textwidth]{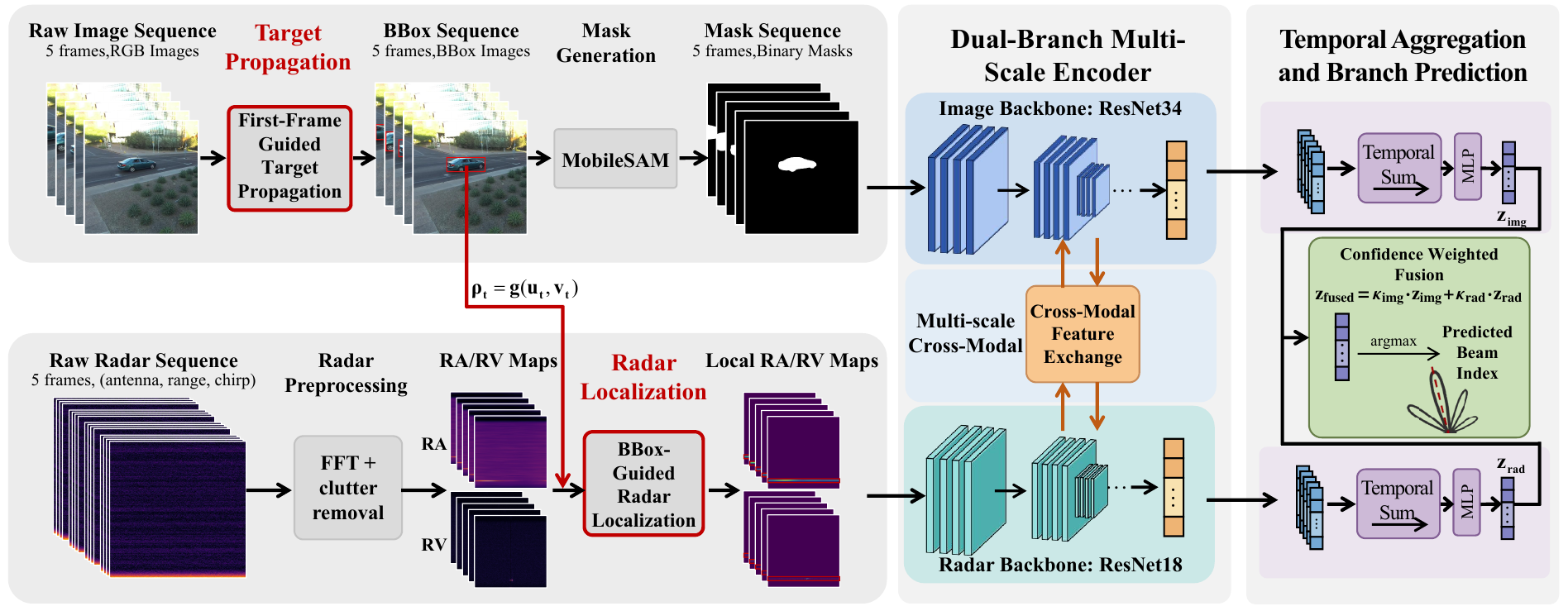}
\caption{Overall target-conditioned beam prediction pipeline. The upper path tracks the visual target, while calibrated image-to-range mapping applies the same association to local RA/RV maps in the lower path. Multi-scale features are exchanged and aggregated over five frames before confidence-weighted fusion.}
\label{fig:model_new}
\end{figure*}

The framework targets crowded, fixed-site V2I scenes and contains three stages (Fig.~\ref{fig:model_new}). First, image-side target propagation follows the target vehicle from a first-frame anchor. Second, radar range localization maps each propagated box to the target's radar range rows. Third, a multimodal network predicts the beam from the target masks and local RA/RV windows.

\subsection{First-Frame Guided Target Propagation}
\begin{figure}[t]
\centering
\includegraphics[width=0.74\columnwidth]{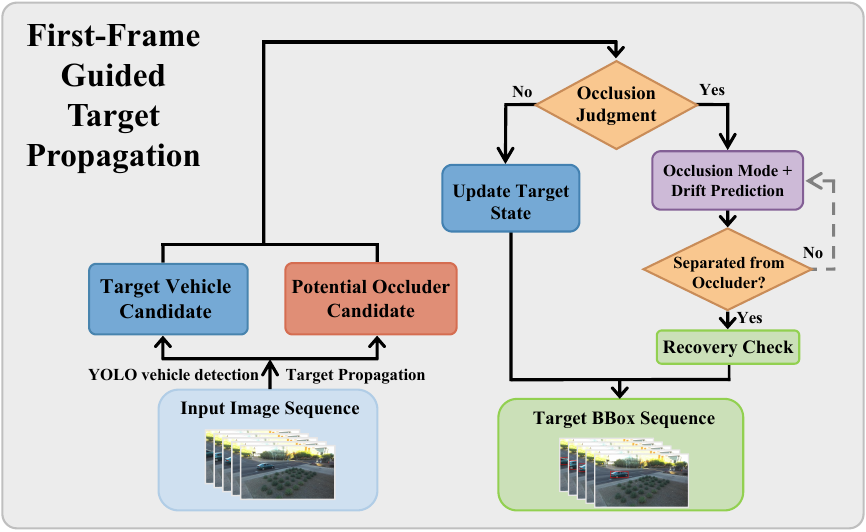}
\caption{First-frame guided target propagation with candidate generation, occlusion judgment, prediction-only fallback, and recovery checking.}
\label{fig:bbox_strategy_new}
\end{figure}

The propagation module (Fig.~\ref{fig:bbox_strategy_new}) updates a single target hypothesis through prediction, candidate scoring, occlusion handling, and recovery. Its target state is
\begin{equation}
\boldsymbol{\xi}_t = \left(\mathbf{b}_t, \mathbf{m}_t, \boldsymbol{\nu}_t, \phi_t\right),
\end{equation}
where $\mathbf{b}_t$ is the target box, $\mathbf{m}_t\in\{0,1\}^{V\times U}$ is its binary mask, $\boldsymbol{\nu}_t$ is the Kalman motion state, and $\phi_t$ is the propagation status. It is \emph{tracked} when a target detection is accepted and no other vehicle is nearby, \emph{occlusion warning} when a detection is accepted but another vehicle approaches or overlaps the box, \emph{occluded} when no detection is accepted while another vehicle overlaps the box, \emph{predicted} when the detection is missed without such overlap, \emph{recovery pending} while a reappearing candidate is confirmed, and \emph{recovered} in the first confirmed frame, after which it returns to tracked.

Propagation starts from the first-frame target box of each segment, which is selected manually in the experiments; in deployment, user-identification or multi-candidate methods can link a decoded user or position cue to the vehicle box \cite{Charan2024UserIdentification,Charan2025MultiCandidate}. The motion prior is taken from Lucas--Kanade flow \cite{LucasKanade1981} on target points when the flow is valid and from local template matching around the previous box otherwise; when both image cues are unreliable, the box is extrapolated with the Kalman-filtered constant-velocity estimate \cite{Kalman1960}, which is also used during fallback in Section~III-C.

In each frame, YOLO11n generates vehicle candidates \cite{UltralyticsYOLO11}, among which the target is selected by a consistency score. For a candidate box $\mathbf b$, let $O_p=\mathrm{IoU}(\mathbf b,\mathbf b_t^{\mathrm{pred}})$, $O_{t-1}=\mathrm{IoU}(\mathbf b,\mathbf b_{t-1})$, and $O_g=\max\{O_p,\mathrm{IoU}(\mathbf b,\bar{\mathbf b}_t)\}$, where IoU denotes the intersection over union, $\mathbf b_t^{\mathrm{pred}}$ is the predicted box, and $\bar{\mathbf b}_t$ is the last reliable target box. With center distance $d$, mean-color distance $\delta_c$, appearance-feature distance $\delta_a$, and relative aspect-ratio change $\delta_q$, the normalized consistencies are
\begin{gather*}
D=[1-d/d_0]_+,\quad C=[1-\delta_c/\tau_c]_+,\\
A=[1-\delta_a/\tau_a]_+,\quad Q=1-\min(1,\delta_q),
\end{gather*}
where $[x]_+=\max(x,0)$ and $d_0$, $\tau_c$, and $\tau_a$ are normalization scales. The color distance compares the mean BGR value of the candidate box with that of the target box in the previous frame, whereas the appearance distance compares a normalized HSV histogram of the box, concatenated with its mean BGR value, with that of $\bar{\mathbf b}_t$ and thus captures the color distribution. With $p_d$ the detector confidence,
\begin{equation}
S_t(\mathbf b)=\boldsymbol\lambda^{\mathsf T}
 [O_p,O_{t-1},D,C,A,Q,O_g,p_d]^{\mathsf T},
\label{eq:candidate_score}
\end{equation}
where $\boldsymbol\lambda$ is a non-negative weight vector and $\hat{\mathbf b}_t=\arg\max_{\mathbf b\in\mathcal G_t}S_t(\mathbf b)$. The admissible set $\mathcal G_t$ applies road-region, displacement, scale, aspect-ratio, color, and appearance gates before ranking; a detection is accepted whenever $\mathcal G_t\neq\emptyset$, except after an occluded or predicted frame, when it must pass the recovery check of Section~III-C. $O_p$ and $O_{t-1}$ favor temporal continuity, while the color and appearance terms separate adjacent vehicles of similar size. The term $O_g$ favors candidates near the last reliable position after a fallback period. MobileSAM \cite{MobileSAM2023} produces the binary mask $\mathbf m_t$ only after the target candidate is accepted, using the accepted box as its prompt. Resized to $256\times256$ by nearest-neighbor interpolation, this mask is the image-branch input $\mathcal M_t$; no RGB values are passed to the predictor.

\subsection{Occlusion Fallback and Recovery}
Overlap with nearby vehicles can cause missed detections or identity switches. Let $\mathbf{c}_t=[u_t,v_t]^{\mathsf{T}}$ denote the target center in image coordinates, and let $\Delta\mathbf{c}^{\mathrm{kf}}_{t-1}\in\mathbb{R}^2$ denote the Kalman-filtered center displacement from frame $t-1$ to frame $t$. In the occluded and predicted states, the target center is updated by
\begin{equation}
\mathbf{c}_t^{\text{pred}} = \mathbf{c}_{t-1} + \Delta\mathbf{c}^{\mathrm{kf}}_{t-1}.
\label{eq:fallback_center}
\end{equation}
The predicted box scale is inherited from recent history with limited variation to prevent drastic drift during short occlusion periods.

When a non-target vehicle first overlaps the selected or predicted target box, it is locked as the occluder and tracked separately. The target stays in the warning state while its detection is still accepted and enters the occluded state once it is not, in which its center is propagated by \eqref{eq:fallback_center}. In the occluded and predicted states, the last reliable mask is cropped and rescaled into the predicted box, so that the mask moves with the box and keeps the pre-occlusion silhouette. The appearance model of the target is frozen so that it does not adapt to the occluder. After an occluded or predicted frame, a candidate near the predicted box is accepted only if its candidate score, appearance similarity, detector confidence, and box overlap meet their thresholds over consecutive frames; otherwise, propagation continues and recovery is retried.

\subsection{Calibrated Image-to-Range Mapping}
The radar observation $\mathcal R_t$ consists of an RA map and an RV map computed from the range fast Fourier transform (FFT) of the raw radar data (Table~\ref{tab:train_config}). Static clutter is removed from the RA map by subtracting the mean over chirps, whereas in the RV map static returns fall in the zero-Doppler column. Since both maps share this range axis and contain the returns of all vehicles in the scene, the target is localized on the range axis from its propagated box. Let $\mathbf q_t=[u_t,v_t,1]^{\mathsf T}$ denote the homogeneous image coordinate of the box center. For a fixed camera--radar installation and a locally planar road region $l$, e.g., a traffic lane, with $l$ determined from the first-frame box, the box centers of vehicles are assumed to lie approximately on a plane parallel to the road at a common effective height. Image points and points on this anchor plane are then related by a homography $\mathbf H_l\in\mathbb R^{3\times3}$, which maps image coordinates directly to plane coordinates centered at the vertical projection of the radar and thus subsumes the camera intrinsics, the camera--radar pose, and the plane geometry. With $\mathbf h_{k,l}^{\mathsf T}$ denoting the $k$th row of $\mathbf H_l$, the planar position of the target is $\mathbf p_{t,l}=[\mathbf h_{1,l}^{\mathsf T}\mathbf q_t,\,\mathbf h_{2,l}^{\mathsf T}\mathbf q_t]^{\mathsf T}/(\mathbf h_{3,l}^{\mathsf T}\mathbf q_t)$. Neglecting the height difference between the radar and the anchor plane, the target range is $\|\mathbf p_{t,l}\|$, and the continuous range row is the following function of image position:
\begin{equation}
\rho^{\mathrm{phys}}_{t,l}=\frac{\|\mathbf p_{t,l}\|-r_{\min}}{\Delta r},
\label{eq:physical_composite_mapping}
\end{equation}
where $r_{\min}$ is the effective range origin and $\Delta r$ is the range-bin spacing. Instead of registering $\mathbf H_l$, $r_{\min}$, and fixed response offsets analytically, we fit this relation directly, which also absorbs any systematic displacement between the box center and the dominant radar response.

\begin{figure}[t]
\centering
\includegraphics[width=0.76\columnwidth]{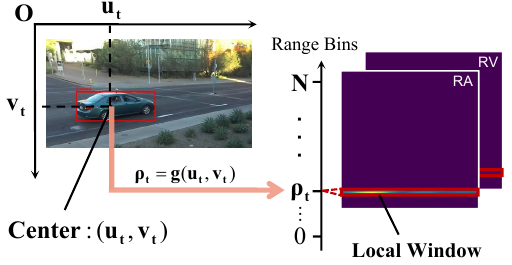}
\caption{Calibration-based radar-domain association. The propagated visual target is mapped to a radar range row, around which a local RA/RV window is retained.}
\label{fig:bbox_x_to_rangebin_new}
\end{figure}

Over a narrow, approximately straight road region, the anchor points lie approximately on a line, and the restriction of $\mathbf H_l$ to this line is a one-dimensional projective map: the coordinate along the line is a fractional linear function of image position with denominator $\mathbf h_{3,l}^{\mathsf T}\mathbf q_t$. When the lateral offset of the radar from the line is small compared with the range, the range is approximately an affine function of this coordinate. The range row in \eqref{eq:physical_composite_mapping} is therefore approximately a fractional linear function of the position along the line, with a constant numerator. Since the box centers lie close to the image of this line, after normalizing $\bar u=(u-\mu_{u,l})/\sigma_{u,l}$ and $\bar v=(v-\mu_{v,l})/\sigma_{v,l}$ we write it as
\begin{equation}
\rho^{\mathrm{phys}}_{t,l}\approx g_l(u_t,v_t)
=\alpha_l+\frac{\gamma_l}
{a_{u,l}\bar u_t+a_{v,l}\bar v_t+1},
\label{eq:reciprocal_mapping}
\end{equation}
where $\alpha_l$ absorbs the range-row origin and fixed offsets, $\gamma_l$ sets the range scale, and $a_{u,l}$ and $a_{v,l}$ describe the image-direction dependence; the denominator is the normalized form of $\mathbf h_{3,l}^{\mathsf T}\mathbf q_t$ in $\mathbf p_{t,l}$. The parameters $\widehat{\boldsymbol\vartheta}_l=(\widehat\alpha_l,\widehat\gamma_l,\widehat a_{u,l},\widehat a_{v,l})$ are identified by least squares from synchronized training correspondences $\mathcal C_l=\{(u_i,v_i,\rho_i^{\mathrm{RA}})\}_{i=1}^{N_l}$
\begin{equation}
\widehat{\boldsymbol\vartheta}_l
=\arg\min_{\alpha,\gamma,a_u,a_v}
\sum_{i=1}^{N_l}\left[
\rho_i^{\mathrm{RA}}-\alpha-
\frac{\gamma}{a_u\bar u_i+a_v\bar v_i+1}
\right]^2,
\label{eq:reciprocal_fit}
\end{equation}
with the denominator constrained to remain positive and away from zero over the road region, and $\widehat\rho_{t,l}=g_l(u_t,v_t;\widehat{\boldsymbol\vartheta}_l)$ is used at inference. The correspondences require no camera--radar registration. From the training set, we select continuous segments in which the target is the only vehicle, a few hundred frames in total, and manually verify that the detected box belongs to that vehicle. Because static clutter is removed from the RA map, its dominant return in these frames comes from the target, and the range row of the peak of the angle-summed RA map gives $\rho_i^{\mathrm{RA}}$. The fitted mapping remains fixed on validation and test data. As shown in Fig.~\ref{fig:bbox_x_to_rangebin_new}, the local maps $\widetilde{\mathcal{R}}^{\mathrm{RA}}_t$ and $\widetilde{\mathcal{R}}^{\mathrm{RV}}_t$ retain the rows from $\bar\rho_{t,l}-w$ to $\bar\rho_{t,l}+w$ around $\bar\rho_{t,l}=\operatorname{round}(\widehat\rho_{t,l})$, with all angle/velocity columns, and zero the remaining rows; we use the half-width $w=3$.

\setlength{\textfloatsep}{0.9\baselineskip plus 0.2\baselineskip minus 0.2\baselineskip}

\subsection{Beam Prediction Backbone}
The predictor follows the cross-modal exchange and energy-based fusion of CMDF \cite{Zhu2025CMDF}, with a ResNet34 image branch and a ResNet18 radar branch \cite{He2016ResNet}. Image--radar interaction uses exchange blocks after the second, third, and fourth residual stages. Each block average-pools the image and radar feature maps of all five frames to $8\times8$ grids and passes the resulting tokens, with a learned positional embedding, through an eight-layer transformer with four attention heads. The outputs are bilinearly upsampled and added to the original feature maps, which are then refined by efficient channel attention (ECA) \cite{Wang2020ECA}.

The frame-level features are summed over the five-frame window, projected by branch-specific multilayer perceptrons, and converted to branch logits. The final beam decision uses this confidence-weighted fusion: if $\mathbf{z}^{(m)}_t$ denotes the logit vector of modality branch $m$ and $E^{(m)}_t = -\log \sum_j \exp (z^{(m)}_{t,j})$ its energy, the fused logit is
\begin{equation}
\mathbf{z}^{\text{fused}}_t = \sum_m \kappa^{(m)}_t \, \mathbf{z}^{(m)}_t, \qquad \kappa^{(m)}_t = -0.1 E^{(m)}_t,
\end{equation}
where the coefficient $0.1$ is a fixed scale. Since the weights are not normalized, we verified that $\kappa^{(m)}_t>0$ for every test sample in the five trained models, so neither branch is sign-inverted, and the image branch receives a median share of $0.49$--$0.51$ of the total weight.

\begin{figure*}[t]
\centering
\includegraphics[width=\textwidth]{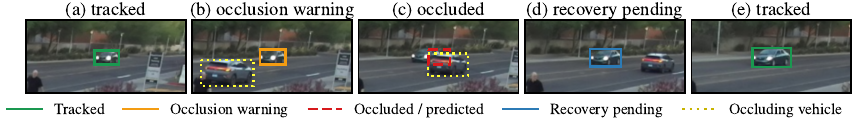}
\caption{Propagation states during an occlusion by an oncoming vehicle: (b) warning as the vehicle approaches, (c) predicted box while the target is hidden and the occluder is tracked, (d) confirmation of the reappearing target, and (e) tracking after recovery.}
\label{fig:tracking_states}
\end{figure*}

\section{Experimental Evaluation}
\subsection{Experimental Setup}
We evaluate the framework on Scenario~31 of the DeepSense 6G dataset \cite{Alkhateeb2023DeepSense}, a roadside V2I deployment with synchronized camera, radar, and 64-beam measurements. Of the test frames, $99.3\%$ contain at least two vehicles and $77.1\%$ at least three. To prevent temporal leakage, the continuous footage is partitioned into non-overlapping segments of at least 50 frames, and each segment is assigned as a whole to the train, validation, or test set, giving 5,336/734/684 windows of $L=5$ frames. The main settings are listed in Table~\ref{tab:train_config}. Every configuration, including the baselines, is trained five times (seeds 100--104) with the checkpoint selected by validation DBA, and the mean test performance is reported. Besides Top-1 accuracy, we report the distance-based accuracy
\begin{equation}
\mathrm{DBA}=\frac{1}{K}\sum_{k=1}^{K}\left(1-\frac{1}{N}\sum_{n=1}^{N}\min_{1\le i\le k}\min\!\left(\frac{|\hat j_{n,i}-j^{\star}_n|}{\Delta_{\mathrm{tol}}},1\right)\right),
\end{equation}
where $\hat j_{n,i}$ is the $i$th-ranked prediction for test sample $n$, $K=3$, and the beam tolerance $\Delta_{\mathrm{tol}}=5$. 

\begin{table}[t]
\caption{Main Experimental Settings}
\label{tab:train_config}
\centering
\footnotesize
\begin{tabular}{ll}
\toprule
Setting & Value \\
\midrule
Window / image size & $5$ frames / $256\times256$ \\
Image / radar backbone & ResNet34 / ResNet18 \cite{He2016ResNet} \\
Loss & one-vs-rest sigmoid focal loss \cite{Lin2017Focal} \\
Focal parameters & focusing $2$, balancing $0.25$ \\
Optimizer & AdamW \cite{Loshchilov2019AdamW}, lr = weight decay = $10^{-4}$ \\
Batch size / epochs & 24 / 50 \\
RA / RV bins & $256$ range $\times$ $256$ angle / Doppler \\
RA / RV sum & over chirps / antennas, min--max normalized \\
\bottomrule
\end{tabular}
\end{table}

\begin{table}[t]
\caption{Beam Prediction on DeepSense 6G Scenario 31}
\label{tab:main_results}
\centering
\footnotesize
\setlength{\tabcolsep}{4pt}
\begin{tabular}{llcc}
\toprule
Method & Image / radar input & DBA & Top-1 (\%) \\
\midrule
TII \cite{Tian2023TII} & scene RGB / global & 0.8299 & 44.97 \\
DEQFusion \cite{Ni2023DEQ} & scene RGB / global & 0.8344 & 44.39 \\
CMDF \cite{Zhu2025CMDF} & scene RGB / global & 0.8473 & 47.95 \\
\midrule
Scene-level baseline & scene RGB / global & 0.8463 & 47.69 \\
+ radar conditioning & scene RGB / local & 0.8787 & 48.83 \\
+ visual conditioning & target mask / global & 0.8788 & 48.83 \\
Proposed & target mask / local & \textbf{0.8821} & \textbf{51.66} \\
\bottomrule
\end{tabular}
\end{table}

\subsection{Comparison with Image--Radar Fusion Methods}
Table~\ref{tab:main_results} compares the proposed framework with the image--radar fusion networks TII \cite{Tian2023TII}, DEQFusion \cite{Ni2023DEQ}, and CMDF \cite{Zhu2025CMDF}, all fed with the scene RGB image and global RA/RV maps. The lower part separates the two forms of target conditioning on the same backbone, where local radar keeps the $2w+1$ rows around $\bar\rho_{t,l}$ and the image input is the binary mask $\mathcal M_t$.

With scene-level inputs, the backbone performs on par with CMDF, whose fusion it follows, and better than TII and DEQFusion. The improvement of the proposed framework therefore comes from target conditioning rather than from the fusion network. Each form of conditioning alone raises DBA by about $0.032$, to $0.8787$ with local radar and $0.8788$ with the target mask. The full method reaches a DBA of $0.8821$ and a Top-1 accuracy of $51.66\%$, which is $2.83\%$ above either mixed variant and $3.71\%$ above CMDF. The local window contributes by removing the returns of other vehicles and static objects at other ranges, which a global RA/RV map presents with equal weight.

\subsection{Target Propagation versus Multi-Object Tracking}
Fig.~\ref{fig:tracking_states} illustrates the propagation states through an occlusion. To evaluate the propagation quantitatively, we replace it with DeepSORT \cite{Wojke2017DeepSORT}, which has been used for vision-aided beam tracking \cite{Li2024VAMUBT}, ByteTrack \cite{Zhang2022ByteTrack}, BoT-SORT \cite{Aharon2022BoTSORT}, OC-SORT \cite{Cao2023OCSORT}, and the nearest-neighbor box association of \cite{Park2025Occlusion}. All of them use the same detector and first-frame anchor, and each box sequence passes through the same mask generation, radar mapping, and network training as the proposed method. For a fair comparison, the parameters of each tracker are tuned on the training set to minimize the miss rate after occlusions. Without manual box labels, tracking is scored against the GPS position of the target, projected into the image by a homography fitted on the manually verified single-vehicle training frames of Section~III-D; a frame is a miss if this point lies outside the box.

\begin{table}[t]
\caption{Tracking Accuracy and Resulting Beam Prediction for Different Box Sources (Target Mask + Local RA/RV)}
\label{tab:trackers}
\centering
\footnotesize
\setlength{\tabcolsep}{3pt}
\begin{tabular}{lcccccc}
\toprule
 & \multicolumn{2}{c}{Miss rate (\%)} & Center & Recov. & & Top-1 \\
\cmidrule(lr){2-3}
Box source & All & Occl. & err. (px) & occl. & DBA & (\%) \\
\midrule
DeepSORT \cite{Wojke2017DeepSORT} & 35.7 & 41.3 & 82.8 & 7/19 & 0.6322 & 32.37 \\
ByteTrack \cite{Zhang2022ByteTrack} & 14.7 & 20.9 & 45.4 & 14/19 & 0.7736 & 42.46 \\
BoT-SORT \cite{Aharon2022BoTSORT} & 16.2 & 21.4 & 47.5 & 15/19 & 0.7816 & 41.17 \\
OC-SORT \cite{Cao2023OCSORT} & 15.5 & 21.8 & 44.4 & 14/19 & 0.7810 & 42.34 \\
Nearest neighbor \cite{Park2025Occlusion} & 23.7 & 50.0 & 78.5 & 4/19 & 0.6652 & 33.33 \\
\midrule
Proposed & \textbf{0.6} & \textbf{5.3} & \textbf{13.0} & \textbf{19/19} & \textbf{0.8821} & \textbf{51.66} \\
\bottomrule
\end{tabular}
\end{table}

Table~\ref{tab:trackers} lists the miss rates over all test frames and over the 21 manually annotated occlusion intervals, the mean center error to the GPS-associated detection during them, the recovered occlusions ($\ge 80\%$ hits in the next 10 frames), and the beam prediction. ByteTrack, BoT-SORT, and OC-SORT miss the target in $14.7$--$16.2\%$ of the test frames and a fifth of the occluded frames, mostly through identity loss rather than a box on another vehicle. When a passing vehicle hides the distant target, the reappearing, partially visible target fails to match its old track and continues under a new identity unrelated to the anchor. DeepSORT loses the target even more often, because its motion gate rejects the reappearing target before the appearance is compared; a vehicle re-identification embedding, public or trained on our training split, does not help and gives $45.6$--$46.1\%$ during occlusions. Nearest-neighbor association never loses a track, but it switches to an adjacent vehicle whenever two boxes meet, which gives the highest miss rate during occlusions. The proposed propagation instead tracks the occluder, freezes the target appearance, and accepts a candidate only after consistent frames. It misses $5.3\%$ of the occluded frames and recovers the target after all 19 occlusions followed by further frames in the same segment.

The tracking errors carry over directly to beam prediction. With the boxes of any of the trackers, the full pipeline falls below the scene-level baseline (DBA $0.6322$--$0.7816$ versus $0.8463$), even though the network is identical to that of the proposed method. Target conditioning removes the evidence outside the box: the mask is zero elsewhere, and only $2w+1$ radar rows are kept. A scene-level input still contains the target vehicle, which the network can learn to locate, whereas a misplaced box leaves no evidence of the target at all. Consistently, on test windows where a tracker's box covers the target in all five frames, its DBA reaches $0.850$--$0.873$, close to the full method, but on the remaining $16$--$39\%$ of the windows it drops to $0.152$--$0.369$. Only $1.2\%$ of the windows are off target with the proposed propagation. Target conditioning thus pays off only when the first-frame identity survives occlusion.

\section{Conclusions}
This paper links the target vehicle across camera and radar data for crowded V2I beam prediction. The first-frame guided propagation maintains the target identity through occlusions and loses the target in only $0.6\%$ of the test frames on Scenario~31, substantially fewer than existing multi-object trackers, and the calibrated image-to-range mapping transfers this identity to local RA/RV windows. Joint visual and radar target conditioning raises DBA from $0.8463$ to $0.8821$ and Top-1 accuracy from $47.69\%$ to $51.66\%$ over the scene-level baseline, and outperforms three image--radar fusion networks.

\vspace{-4pt}
\section*{Acknowledgment}
\vspace{-2pt}
This work was supported in part by National Natural Science Foundation of China under Grant U22B2057, and in part by State Key Laboratory of Environment Characteristics and Effects for Near-space under Grant 2025NS10. (Corresponding authors: Zhaocheng Wang and Xiao Liang.)

\end{document}